# Chapter 2

# Future Climate Change Projections over the Indian Region

**Lead Author:** *J. Sanjay*
**Co-authors:** *R. Krishnan, M.V.S. Ramarao, R. Mahesh, Bhupendra Singh, Jayashri Patel, Sandip Ingle, Preethi Bhaskar, J.V. Revadekar, T.P. Sabin, M. Mujumdar*

# 1. Introduction

Assessments of impacts of climate change and future projections over the Indian region, have so far relied on a single regional climate model (RCM) - eg., the PRECIS RCM of the Hadley Centre, UK. While these assessments have provided inputs to various reports (e.g., INCCA 2010; NATCOMM2 2012), it is important to have an ensemble of climate projections drawn from multiple RCMs due to large uncertainties in regional-scale climate projections. Ensembles of multi-RCM projections driven under different perceivable socio-economic scenarios are required to capture the probable path of growth, and provide the behavior of future climate and impacts on various biophysical systems and economic sectors dependent on such systems.

The Centre for Climate Change Research, Indian Institute of Tropical Meteorology (CCCR-IITM) has generated an ensemble of high resolution downscaled projections of regional climate and monsoon over South Asia until 2100 for the Intergovernmental Panel for Climate Change (IPCC) using a RCM (ICTP-RegCM4) at 50 km horizontal resolution, by driving the regional model with lateral and lower boundary conditions from multiple global atmosphere-ocean coupled models from the Coupled Model Intercomparison Project Phase 5 (CMIP5). The future projections are based on three Representation Concentration Pathway (RCP) scenarios (viz., RCP2.6, RCP4.5, RCP8.5) of the IPCC.

These high-resolution downscaled projections of regional climate over South Asia are generated as part of the International Programme called Coordinated Regional Downscaling Experiment (CORDEX) sponsored by the World Climate Research Programme. This chapter provides a synthesis of results from the CORDEX South Asia multi-RCM outputs, that allows us to interpret the strengths and limitations of future regional climate projections over India. This information is useful to reduce uncertainty of impact assessment estimates to an extent and provide a pan-Indian regional assessment for informed policy-making.

## Highlights

- The all India mean surface air temperature change for the near-term period 2016–2045 relative to 1976–2005 is projected to be in the range of 1.08°C to 1.44°C, and is larger than the natural internal variability. This assessment is based on a reliability ensemble average (REA) estimate incorporating each RCM performance and convergence, and is associated with less than 16% uncertainty range (Table 2.1, Box 2.4).
- The all India mean surface air temperature is projected to increase in the far future (2066–2095) by 1.35 ± 0.23°C under RCP2.6, 2.41 ± 0.40°C under RCP4.5 and 4.19 ± 0.46°C under RCP8.5 scenario respectively. These changes are relative to the period 1976–2005. The semi-arid north-west and north India will likely warm more rapidly than the all India mean (Table 2.1, Fig. 2.1).
- Monthly increase in all India mean surface air temperature based on REA estimate is relatively higher during winter months than in the summer monsoon months throughout the 21$^{st}$ century under the three RCP scenarios (Fig. 2.3).
- The REA changes for all India annual minimum temperature of 4.43 ± 0.34°C is more pronounced than that of 3.94 ± 0.45°C and 4.19 ± 0.46°C increases estimated for the respective annual maximum and mean temperatures respectively the end of the 21st century under RCP8.5 scenario (Tables 2.1, 2.2 and 2.3).
- The models project substantial changes in temperature extremes over India by the end of the 21$^{st}$ century, with a likely overall decrease in the number of cold days and nights, and increase in the number of warm days and nights.
- Although the all India annual precipitation is found to increase as temperature increases, the REA assessment indicates that precipitation changes throughout the 21$^{st}$ century remain highly uncertain.
- The all India annual precipitation extremes are projected to increase with relatively higher uncertainty under RCP8.5 scenario by the end of the 21$^{st}$ century.
- The downscaled projections suggest that intensification of both dry and wet seasons is expected along the west coast of India and in the adjoining peninsular region.

## 2. Future Projections of Climate over India

The focus of this chapter is on the summary of new and emerging knowledge since the IPCC AR5(IPCC 2013, 2014), with emphasis on material deriving from dynamical downscaling work under CORDEX South Asia (Sanjay et al., 2017), which is often of greater relevance for impact, adaptation and vulnerability (IAV) applications than the coarser resolution CMIP5 global climate model data used for AR5. The assessed downscaled historical and future projections of climate change till the end of the 21$^{st}$ century are based on six simulations with IITM-RegCM4 RCM and ten simulations with SMHI-RCA4 RCM for RCP4.5 and RCP8.5 scenarios (see list of CORDEX South Asia experiments in Table 2.1). While for RCP2.6 only a subset of available five simulations with SMHI-RCA4 runs are assessed. The changes are considered over the Indian land mass, by masking out the oceans and territories outside the geographical borders of India, and are expressed relative to the reference base line period: 1976–2005. The multi-RCM ensemble mean spatial patterns and all India averages are reported for the 30-year future periods: 2016–2045, 2036-2065 and 2066-2095 representing near-term, mid-term and long-term changes in future climate over India.

### 2.1 Projected Changes in Temperature

The projections of near-term change in the CORDEX South Asia multi-RCM ensemble mean annual mean surface air temperature relative to the reference period 1976-2005, show modest sensitivity to alternate RCP scenarios over Indian land area (see left panels of Fig. 2.1). The RCP2.6 scenario shows increase of less than 1°C over most of India except in some areas (eg.,Tamil Nadu in south and Jammu and Kashmir in north), where decreases of less than 1°C near-term change in surface temperature are projected. While under the RCP4.5 and RCP8.5 scenarios, the near-term changeshow similar increase of less than 2°C uniformly over the Indian land. The long-term projections of surface air temperature change over India for the end of 21$^{st}$ century are found to be dependent on the RCP scenarios (see right panels of Fig. 2.1). The geographical patterns of change for RCP2.6 scenario remain below 2°C above 1976-2005 levels over most parts of India throughout the 21st century, where the forced signal is typically smaller than the internal variability of the climate system (see top panels of Fig. 2.1). The multi-RCM ensemble mean annual mean surface air temperature mid-term change exceeds 2°C over the north-west and north India for RCP4.5 scenario, while in the long-term the change for this mid-scenario exceeds 2°C over most parts of India except the southern tip of the Indian peninsula (see middlepanels of Fig. 2.1). The spatial pattern and magnitude of the projected mid-term warming for the RCP8.5 scenario resembles that of the long-term change for the RCP4.5 scenario (see bottompanels of Fig. 2.1). The projected annual warmingexceeding 3°C over entire India is more rapid for this high-emission scenario by the end of 21st century, with relatively higher change exceeding 4°C projected in the semi-arid north-west and north India (see bottom right panel of Fig. 2.1).

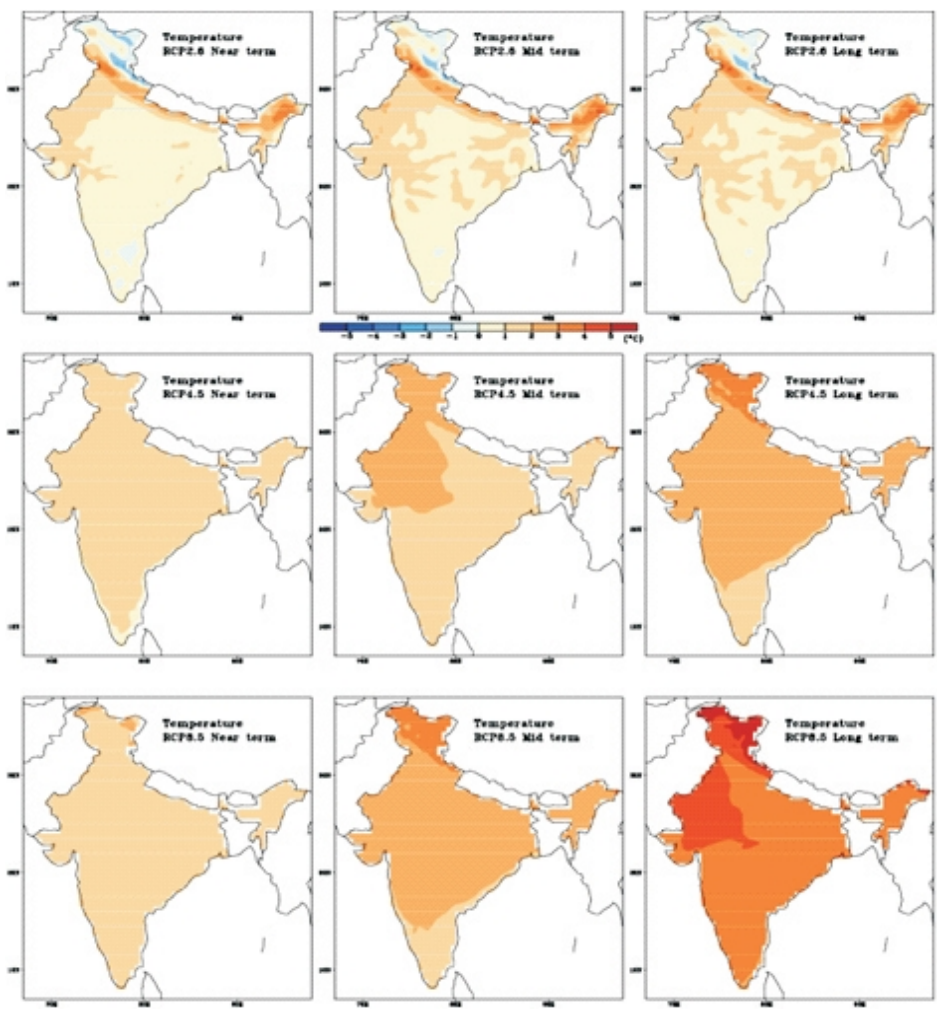

*Figure 2.1 CORDEX South Asia multi-RCM ensemble mean projections of annual average surface air temperature (°C) changes for near-term (2016-2045), mid-term (2036–2065) and long-term (2066–2095) climate under RCP2.6, RCP4.5 and RCP8.5 scenarios, relative to 1976–2005.*

**Table 2.1** List of the 16 CORDEX South Asia RCM simulations driven with 10 CMIP5 AOGCMs.

| CORDEX South Asia RCM | RCM Description | Contributing CORDEX Modeling Center | Driving CMIP5 AOGCM (see details at https://verc.enes.org/data/enes-model-data/cmip5/resolution) | Contributing CMIP5 Modeling Center |
|---|---|---|---|---|
| IITM-RegCM4 (6 members) | The Abdus Salam International Centre for Theoretical Physics (ICTP) Regional Climatic Model version 4 (RegCM4; Giorgi et al., 2012) | Centre for Climate Change Research (CCCR), Indian Institute of Tropical Meteorology (IITM), India | CCCma-CanESM2 | Canadian Centre for Climate Modelling and Analysis (CCCma), Canada |
| | | | NOAA-GFDL-GFDL-ESM2M | National Oceanic and Atmospheric Administration (NOAA), Geophysical Fluid Dynamics Laboratory (GFDL), USA |
| | | | CNRM-CM5 | Centre National de RecherchesMétéorologiques (CNRM), France |
| | | | MPI-ESM-MR | Max Planck Institute for Meteorology (MPI-M), Germany |
| | | | IPSL-CM5A-LR | Institut Pierre-Simon Laplace (IPSL), France |
| | | | CSIRO-Mk3.6 | Commonwealth Scientific and Industrial Research Organization (CSIRO), Australia |
| SMHI-RCA4 (10 members) | Rossby Centre regional atmospheric model version 4 (RCA4; Samuelsson et al., 2011) | Rosssy Centre, Swedish Meteorological and Hydrological Institute (SMHI), Sweden | ICHEC-EC-EARTH | Irish Centre for High-End Computing (ICHEC), European Consortium (EC) |
| | | | MIROC-MIROC5 | Model for Interdisciplinary Research On Climate (MIROC), Japan Agency for Marine-Earth Sci. & Tech., Japan |
| | | | NCC-NorESM1 | Norwegian Climate Centre (NCC), Norway |
| | | | MOHC-HadGEM2-ES | Met Office Hadley Centre for Climate Change (MOHC), United Kingdom |
| | | | CCCma-CanESM2 | CCCma, Canada |
| | | | NOAA-GFDL-GFDL-ESM2M | NOAA, GFDL, USA |
| | | | CNRM-CM5 | CNRM, France |
| | | | MPI-ESM-LR | MPI-M, Germany |
| | | | IPSL-CM5A-MR | IPSL, France |
| | | | CSIRO-Mk3.6 | CSIRO, Australia |

The all India averaged annual surface air temperature anomalies(relative to 1976–2005) based on the India Meteorological Department (IMD) 1°longitude-latitude gridded data show steady long-term warmingwith interannual variations for the period 1970-2015 (Fig. 2.6). The CORDEX South Asia historical RCM simulations capture the observed the interannualvariationsand the warming trend reasonably well.A consistent and robust feature across the downscaled CORDEX South Asia RCMs is a continuation of warming over India in the 21st century for all the RCP scenarios (Fig. 2.2).The all India averaged annual surface air temperature increases are almost the same for all the RCPscenarios during the first decade after 2005.The warming rate depend more on the specified green house gas concentration pathway at longer time scales, particularly after about 2050. The multi-RCM ensemble mean under RCP2.6 scenario stays around 1.5°C above 1976-2005 levels throughout the 21st century, clearly demonstrating the potentialof mitigation policies.The ensemble mean annual India warming exceeds 2°C within the 21st century under RCP4.5, and the warming exceeds 4°C by the end of the 21st century under RCP8.5 scenario.The spread in the minimum to maximum range in the projected warming among the CORDEX South Asia RCMs for each RCP scenario (shown as shading in Fig. 2.2) provide a simple, but crude, measure of uncertainty.

A reliability ensemble averaging (REA) technique is used to provide a quantitative estimate of the associated uncertainty range of future climate change projections for India simulated by the RCMs under CORDEX South Asia.The viability of REA methodology in providing realistic future CMIP5 projections of the Indian summer monsoon by incorporating model performance and model convergence criteria was demonstrated by Sengupta and Rajeevan (2013) for two main variables, surface air temperature and precipitation. The results of applying REA methodology to the CORDEX South Asia multi-RCMs all India averaged annual surface air temperature changes under the three different RCP scenarios are summarized in Table 2.2.

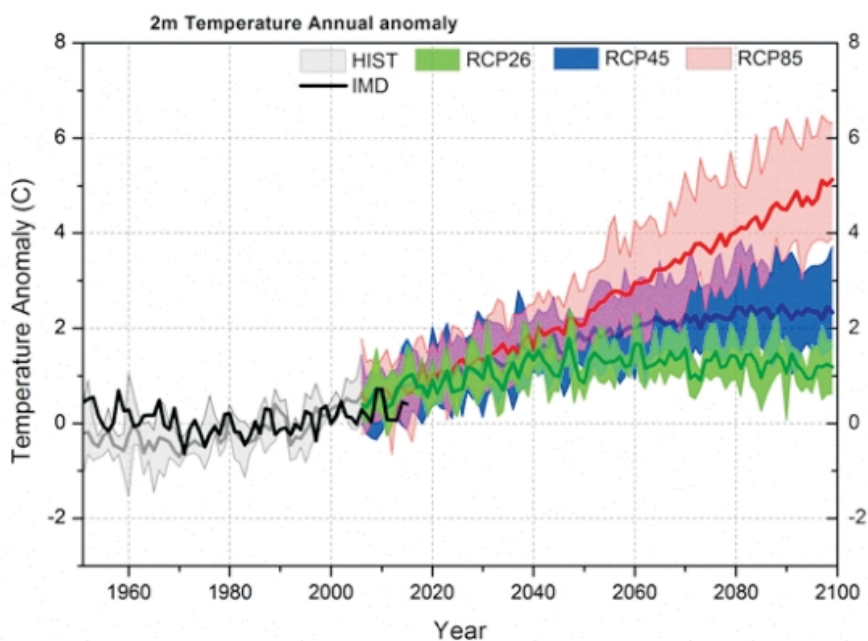

*Figure 2.2* Time series of Indian annual mean surface air temperature (°C) anomalies (relative to 1976–2005) from CORDEX South Asia concentration-driven experiments. The historical simulations (grey) and the downscaled projections are shown for RCP2.6 (green), RCP4.5 (blue) and RCP8.5 (red) scenarios for the multi-RCM ensemble mean (solid lines) and the minimum to maximum range of the individual RCMs (shading). The black line shows the observed anomalies during 1951-2015based on IMD gridded data.

**Table 2.2** CORDEX South Asia multi-RCM reliability ensemble average (REA) estimate of projected changes in annual mean surface air temperature over India and the associated uncertainty range. The values in parenthesis show the uncertainty in percent for the REA estimate.

| Scenario | Annual Mean Temperature (°C) | | |
|---|---|---|---|
| | 2030s | 2050s | 2080s |
| RCP2.6 | 1.08 ± 0.12 *(11.1%)* | 1.35 ± 0.18 *(13.3%)* | 1.35 ± 0.23 *(17.0%)* |
| RCP4.5 | 1.28 ± 0.20 *(15.6%)* | 1.92 ± 0.28 *(14.6%)* | 2.41 ± 0.40 *(16.6%)* |
| RCP8.5 | 1.44 ± 0.17 *(11.8%)* | 2.41 ± 0.28 *(11.6%)* | 4.19 ± 0.46 *(11.0%)* |

The REA near-term (2016-2045) warming are almost the same for all the RCP scenarios ranging between 1.08-1.44°C. The natural variability (computed following Sengupta and Rajeevan (2013)) in the observed all India annual mean surface air temperature (based on IITM India averaged monthly data) is 0.347°C, while the REA based temperature increases are well above this natural variability estimate. The uncertainty range defined by the root mean square difference varies in the near-term between 0.12°C to 0.20°C for the three RCP scenarios, with the RCP4.5 near-term warming of 1.28°C indicating the maximum uncertainty of 15.6% . The proper weighting of individual CORDEX South Asia RCMs based on their present day performance by the REA method has resulted in the REA warming under RCP2.6 scenario to be 1.35°C above 1976-2005 levels till the end of 21st century, which is lesser than that defined by multi-RCM ensemble mean shown in Fig. 2.2. However this estimate of annual warming well below 2°C by the end of 21st century for RCP2.6 is found to be associated with the highest uncertainty of 17% among all the RCP scenarios. The REA estimates of long-term (2066-2095) warming are 2.41±0.40°C and 4.19 ± 0.46°C under RCP4.5 and RCP8.5 scenarios respectively. The assessment of 4.19°C warming by the end of the 21st century under RCP8.5 scenario is highly reliable as it is associated with the lowest uncertainty (of 11%) among the three RCP scenarios.

The REA estimate of projected monthly changes for the CORDEX South Asia multi-RCMs of all India averaged monthly surface air temperature relative to the reference period 1976-2005 indicate relatively higher seasonal warming during winter months than in the summer monsoon months (Fig. 2.3). The annual cycle of the near-term change show least sensitivity to alternate RCP scenarios over Indian land area (see left panels of Fig. 2.3).The long-term monthly change over India for the end of 21st century show distinct annual cycle, particularly for the RCP8.5 scenario (see right panels of Fig. 2.3). The REA monthly change estimates are higher (lower) than the annual values (Table 2.2) in January (July) for the mid-term and long-term periods under the three RCP scenarios. The associated uncertainty range for the REA monthly changes (shading in Fig. 2.3) is seen to steadily increase from near-term to long-term, with the highest values for each month by the end of the 21st century under RCP8.5 scenario.

The forced signal of warming occurs not only in the annual mean of daily mean surface air temperature, but also in the annual means of daily maximum and daily minimum temperatures. The results of applying REA technique to the CORDEX South Asia multi-RCMs all India averaged annual means of daily maximum and daily minimum surface air temperature changes under the three different RCP scenarios are summarized in Table 2.3 and Table 2.4 respectively.

**Table 2.3** CORDEX South Asia multi-RCM reliability ensemble average (REA) estimate of projected changes in annual mean of daily maximum temperature over India and the associated uncertainty range. The values in parenthesis show the uncertainty in percent for the REA estimate.

| Scenario | Annual Maximum Temperature (°C) | | |
|---|---|---|---|
| | 2030s | 2050s | 2080s |
| RCP2.6 | 0.99 ± 0.11 *(11.1%)* | 1.26 ± 0.16 *(12.7%)* | 1.27 ± 0.20 *(15.7%)* |
| RCP4.5 | 1.26 ± 0.20 *(15.9%)* | 1.81 ± 0.27 *(14.9%)* | 2.29 ± 0.36 *(15.7%)* |
| RCP8.5 | 1.36 ± 0.16 *(11.8%)* | 2.30 ± 0.31 *(13.5%)* | 3.94 ± 0.45 *(11.4%)* |

**Table 2.4** CORDEX South Asia multi-RCM reliability ensemble average (REA) estimate of projected changes in annual mean of daily minimum temperature over India and the associated uncertainty range. The values in parenthesis show the uncertainty in percent for the REA estimate.

| Scenario | Annual Minimum Temperature (°C) | | |
|---|---|---|---|
| | 2030s | 2050s | 2080s |
| RCP2.6 | 1.16 ± 0.17 *(14.7%)* | 1.44 ± 0.24 *(16.7%)* | 1.35 ± 0.25 *(18.5%)* |
| RCP4.5 | 1.36 ± 0.18 *(13.2%)* | 2.14 ± 0.28 *(13.1%)* | 2.63 ± 0.38 *(14.4%)* |
| RCP8.5 | 1.50 ± 0.16 *(10.7%)* | 2.60 ± 0.23 *(8.8%)* | 4.43 ± 0.34 *(7.7%)* |

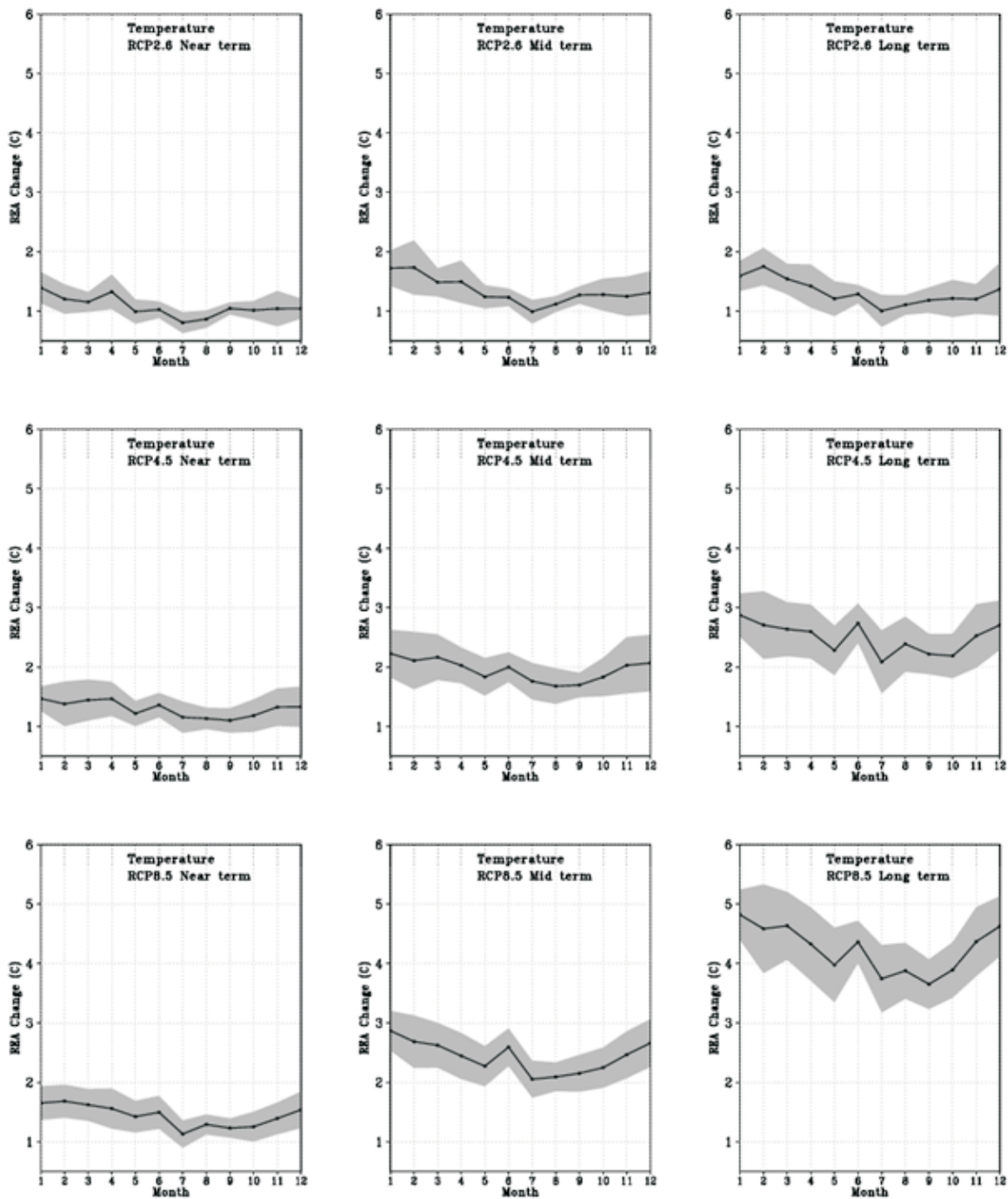

*Figure 2.3* CORDEX South Asia multi-RCM reliability ensemble average (REA) estimate of projected monthly change of all India averaged monthly surface air temperature (°C; solid lines) and the associated uncertainty range (shading) for near-term (2016-2045), mid-term (2036–2065) and long-term (2066–2095) climate under RCP2.6, RCP4.5 and RCP8.5 scenarios, relative to 1976–2005.

The estimate of annual natural variability in the observed all India maximum and minimum surface air temperature (based on IITM India averaged monthly data) are 0.513°C and 0.213°C respectively, while the REA based annual mean maximum and minimum temperature increases are well above these natural variability estimates. The REA estimate of warming for the three 30 year future periods are lower (higher) for the annual means of daily maximum (minimum) temperature than the respective warming found earlier for the annual mean of daily mean temperature under all three RCP scenarios (Table 2.2). The REA changes for annual minimum temperature of 4.43 ± 0.34°C is more pronounced than that of 3.94 ± 0.45°C and 4.19 ± 0.46°C increases estimated for all India annual maximum (Table 2.3) and mean (Table 2.4) temperatures respectively the end of the 21$^{st}$ century under RCP8.5 scenario. The assessment of 4.43°C warming for annual mean of daily minimum surface air temperature by the end of the 21$^{st}$ century under RCP8.5 scenario is highly reliable as it is associated with the lowest uncertainty (of 7.7%) among not only the three RCP scenarios for this variable but also for the annual mean and maximum statistic shown in Table 2.2 and Table 2.3.

Similar to the results obtained for the changes in the all India averaged monthly surface air temperature (Fig. 2.3), the REA estimate of projected monthly changes of maximum and minimum temperature relative to the reference period 1976-2005 indicate relatively higher seasonal warming during winter months than in the summer monsoon months (Fig. 2.4a and 2.4b). The REA estimate of the projected change in the all India minimum temperature for January and December months is likely to exceed 5°C by the end of 21$^{st}$ century under the RCP8.5 scenario (see bottom right panel of Fig. 2.4b).

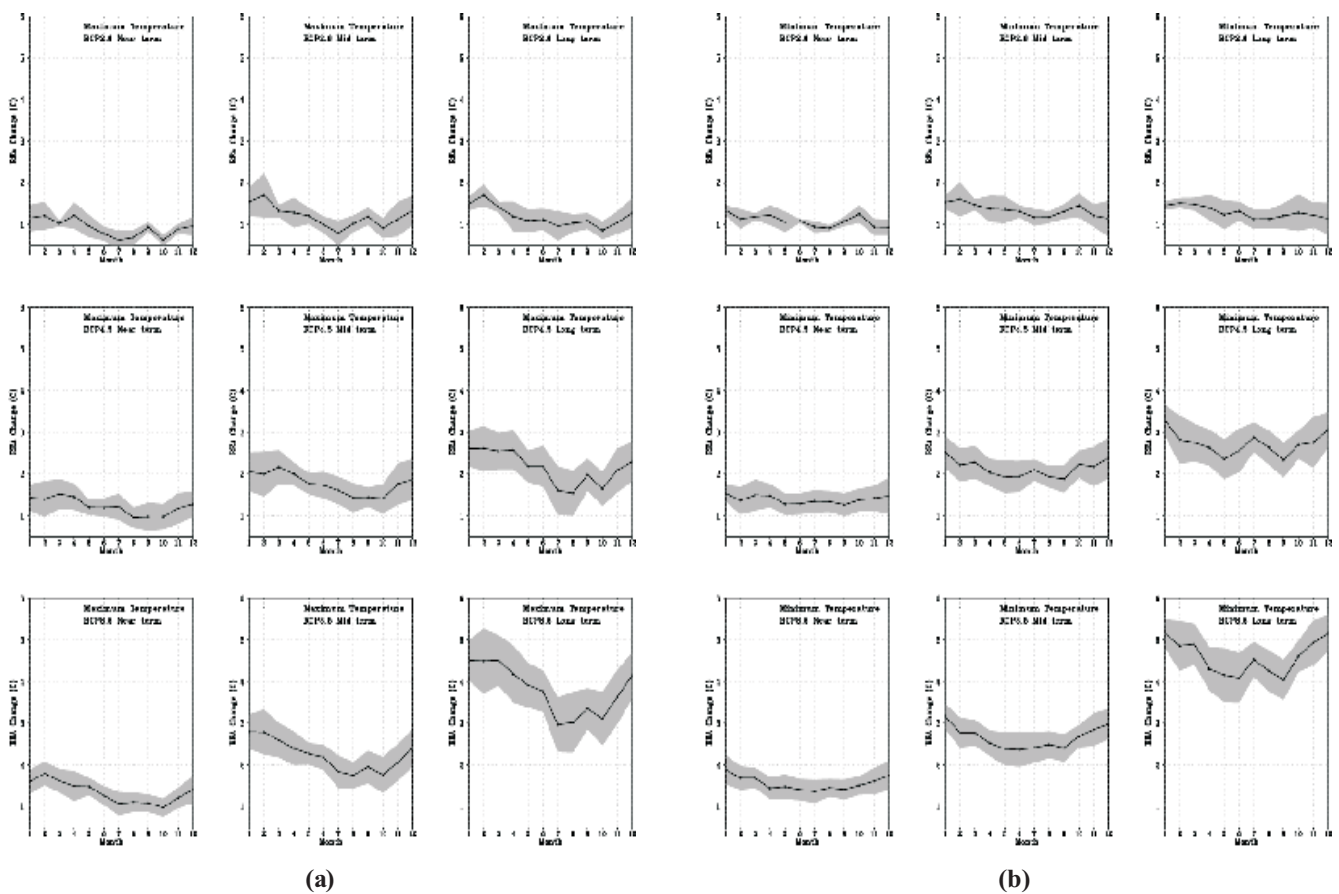

(a)          (b)

*Figure 2.4* CORDEX South Asia multi-RCM reliability ensemble average (REA) estimate of projected monthly change of all India averaged monthly (a) maximum and (b) minimum temperature (°C; solid lines) and the associated uncertainty range (shading) for near-term (2016-2045), mid-term (2036–2065) and long-term (2066–2095) climate under RCP2.6, RCP4.5 and RCP8.5 scenarios, relative to 1976–2005.

## 2.2. Projected Changes in Precipitation

The sign, magnitude, and spatial extent of the projected CORDEX South Asia multi-RCM ensemble mean annual precipitation changes relative to the reference period 1976-2005 exhibit large variability (Fig. 2.5).The dominance of internal variability of the climate system for annual precipitation at sub-regional scales in India over the relatively smaller forced signal under RCP2.6 scenario leads to increasing and decreasing changes for multi-RCM ensemble meanin several parts of the country throughout the 21st century (see top panels of Fig. 2.5). The multi-RCM ensemble mean annual precipitation mid-term increase exceeds 10% over the west coast and the adjoining southern parts of the Indian peninsula for RCP4.5 scenario, while in the long-term the change for this mid-scenario exceeds 20% over the south-west coast and the adjoining Kerala State (see middle panels of Fig. 2.5). The precipitation changes are not significant over the remaining parts of India for this mid-scenario up to the mid 21$^{st}$ century, however in long-term increase exceed 10% over north-west and adjoining parts of the country. The long-term projected annual precipitation increase exceeds 10% over most parts of India except in Jammu and Kashmir under RCP8.5 scenario, with relatively higher increase exceeding 30% projected along the west coast of India for this high-emission scenario by the end of 21$^{st}$ century (see bottom right panel of Fig. 2.5).

The all India averaged annual precipitation anomalies based on IMD high resolution 0.25° longitude-latitude gridded data show interannual variationsduring 1951-2015 (Fig. 2.6). The CORDEX South Asia historical RCM simulations capture the observed interannualvariability. These downscaled projections continue to show interannual variations into the future without any consistent trend for all the RCP scenarios till mid 21$^{st}$ century, and a small increase at longer time scales under RCP8.5 scenario (Fig. 2.6). However the spread in the minimum to maximum range of the projected all India annual precipitation changeamong the CORDEX South Asia RCMs (shown as shading in Fig. 2.6) increases with time, particularly for RCP8.5 scenario, indicating large uncertainty.The results of applying REA methodologyto provide a quantitative estimate of the associated uncertainty range of future climate change projections for Indiaunder the three different RCP scenarios are summarized in Table 2.5.

The natural variability in the observed all India annual mean precipitation (based on IITM India averaged monthly data) is 0.205mm d$^{-1}$, while the REA based precipitation increases are below this natural variability estimate, except in long-term for RCP4.5, and by mid-term for RCP8.5 scenarios. The associated uncertainty range defined by the root mean square difference is also very high for the three RCP scenarios, with the RCP8.5 long-term precipitation increase of 0.55 mm d$^{-1}$ indicating the minimum uncertainty of 58%. Thus the assessment of all India annual precipitation changes throughout the 21$^{st}$ century remains highly uncertain.

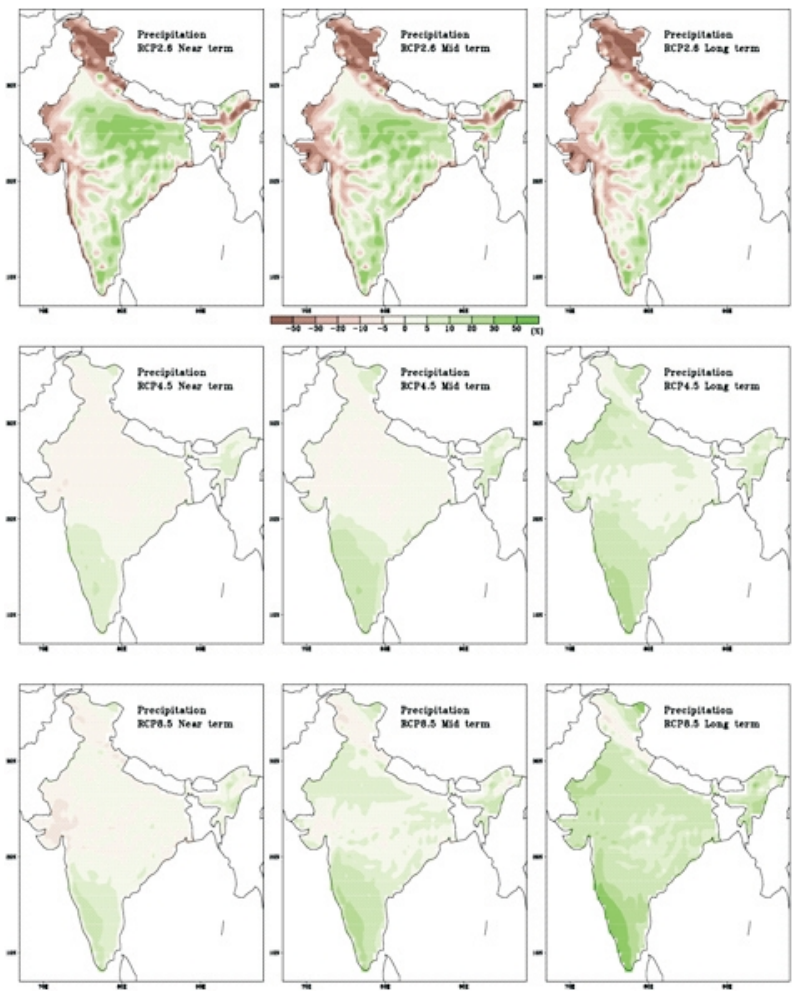

*Figure 2.5* CORDEX South Asia multi-RCM ensemble mean projections of average percent changes in annual mean precipitatio for near-term (2016-2045), mid-term (2036–2065) and long-term (2076–2095) climate under RCP2.6, RCP4.5 and RCP8.5 scenarios, relative to 1976–2005.

**Table 2.5** CORDEX South Asia multi-RCM reliability ensemble average (REA) estimates of projected changes in annual mean precipitation over India and the associated uncertainty range. The values in parenthesis show the uncertainty in percent for the REA estimate.

| Scenario | Annual Mean Precipitation (mm day $^{-1}$) | | |
|---|---|---|---|
| | 2030s | 2050s | 2080s |
| **RCP2.6** | 0.16 ± 0.12 *(75%)* | 0.15 ± 0.17 *(113%)* | 0.14 ± 0.13 *(93%)* |
| **RCP4.5** | 0.07 ± 0.14 *(200%)* | 0.15 ± 0.19 *(127%)* | 0.30 ± 0.21 *(70%)* |
| **RCP8.5** | 0.15 ± 0.15 *(100%)* | 0.27 ± 0.19 *(70%)* | 0.55 ± 0.32 *(58%)* |

The REA estimate of projected monthly changes for the CORDEX South Asia multi-RCMs of all India averaged monthly precipitation relative to the reference period 1976-2005 indicate relatively higher seasonal increase during summer monsoon months than in the winter months (Fig. 2.7). The long-term monthly change over India by the end of 21$^{st}$ century show distinct annual cycle, particularly for the RCP8.5 scenario (see right panels of Fig. 2.7).Although the REA estimate indicates that the projected increase in the all India precipitation during July to October months are expected to exceed 1 mm d$^{-1}$ by the end of the 21$^{st}$ century under RCP8.5 scenario, the associated uncertainty range for the REA monthly changes (shading in Fig. 2.7) are also found to be the highest for these months(see bottom right panel of Fig. 2.7).

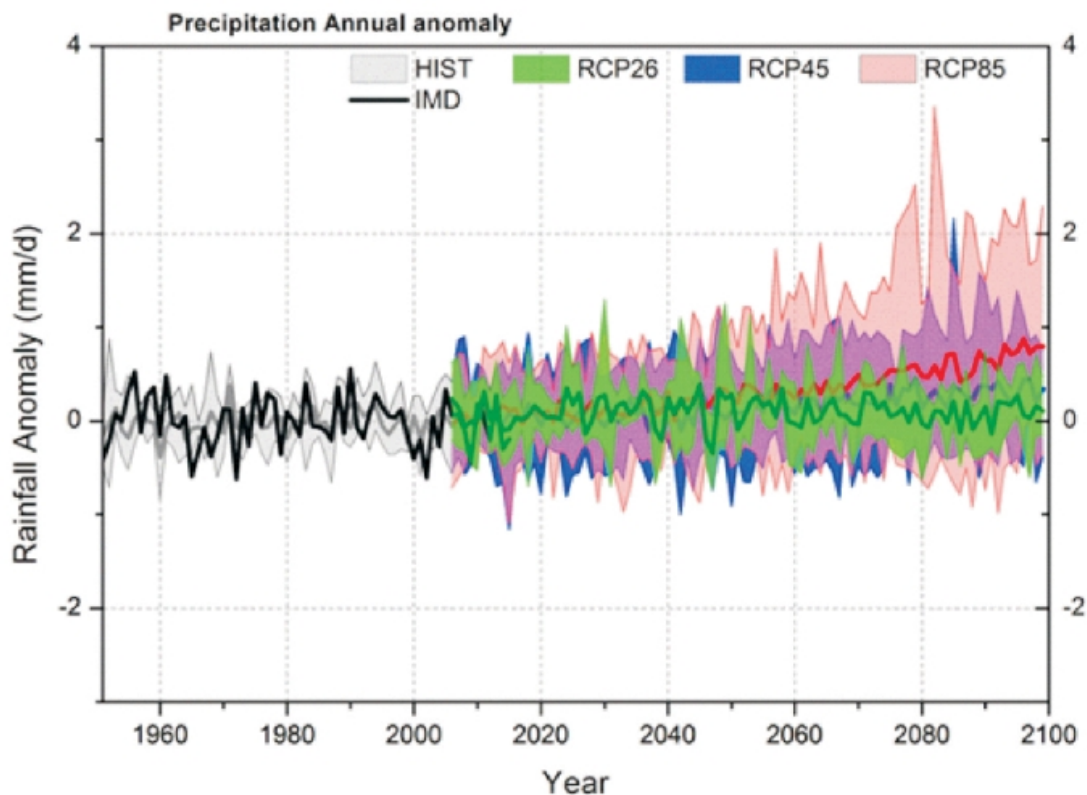

***Figure 2.6*** *Time series of Indian annual mean precipitation (mm d$^{-1}$) anomalies (relative to 1976–2005) from CORDEX South Asia concentration-driven experiments. The historical simulations (grey) and the downscaled projections are shown for RCP2.6 (green), RCP4.5 (blue) and RCP8.5 (red) scenarios for the multi-RCM ensemble mean (solid lines) and the minimum to maximum range of the individual RCMs (shading). The black line shows the observed anomalies during 1951-2015 based on IMD gridded data.*

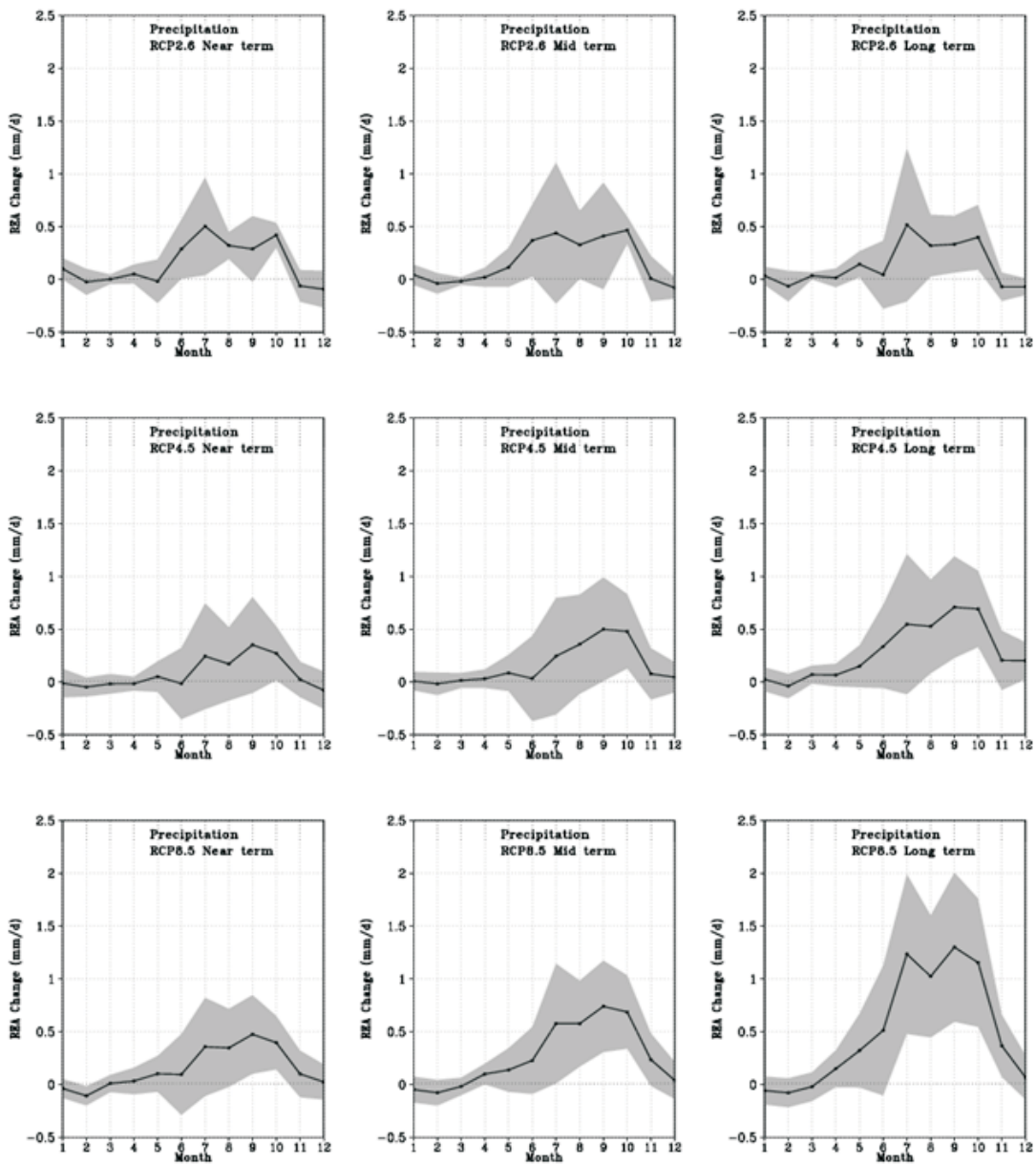

*Figure 2.7* CORDEX South Asia multi-RCM reliability ensemble average (REA) estimate of projected monthly change of all India averaged monthly precipitation (mm d$^{-1}$; solid lines) and the associated uncertainty range (shading)for near-term (2016-2045), mid-term (2036–2065) and long-term (2066–2095) climate under RCP2.6, RCP4.5 and RCP8.5 scenarios, relative to 1976–2005.

## 3. Projected Changes in Climate Extremes over India

The previous section provided information on the projected changes in multi-year averages of annual or monthly climate over India. The climate change vulnerability and impacts result mainly from extreme climate events. The IPCC Special Report on Extremes (SREX; Intergovernmental Panel onClimate Change (IPCC), 2012) assessment concluded that future increases in the number of warm days and nights and decreases in the number of cold days and nights are virtually certain on the global scale.It is also likely that since about 1950 the number of heavy precipitation events over land has increased in more regions than it has decreased (Sillmann et al., 2013).Thissection assessselected temperature- and precipitation-based climate extremes indicesdefined by the Expert Team on Climate Change Detection and Indices (ETCCDI; see details in Box 2.1), which are computed with a consistent methodologyfor climate change simulations using different emission RCP scenarios in the CORDEX South Asia multi-RCM ensemble. The changes in the indices are analysed over Indian land region during 1951 to 2099relative to the reference period 1976–2005.

> **Box 2.1 Temperature and Precipitation Climate Extreme Indices**
>
> *The Expert Team on Climate Change Detection and Indices (ETCCDI) has defined a set of climate change indices to facilitate the investigation of observed and projected changes in temperature and precipitation extremes, focusing on extreme events. These indices in general describe moderate extreme events with a re-occurrence time of 1 year or less (Zhang et al., 2011). The indices are based on daily minimum and maximum of near surface temperature and daily precipitation amounts (TN, TX, and PR, respectively). A subset of the available 27 indices (see Table 1 in Sillmann et al. (2013)) are selected for analysis in this chapter, which give a comprehensive overview of the projected changes in temperature and precipitation extremes across RCMs and scenarios. The calculations are performed with the Climate Data Operators (CDO) package as documented at https://code.mpimet.mpg.de/projects/cdo .*
>
> *The percentile indices for temperature extremes considered are cold nights and days (TN10p and TX10p, respectively) and warm nights and days (TN90p and TX90p, respectively), which describe the threshold exceedance rate of days where TN or TX is below the 10th or above the 90th percentile, respectively. The thresholds are based on the annual cycle of the percentiles calculated for a 5 day sliding window centered on each calendar day in the base period 1976-2005. Sillmann et al. (2013) discussed that these percentile indices are better suited for the tropical regions than absolute temperature indices as they show the highest increase in the tropical regions where inter-annual temperature variability is relatively small. Therefore, small shifts in the mean of the temperature distribution can lead to larger changes in the exceedance rates than in the high variability extra-tropical regions. Since ecosystems and human infrastructure in the tropics are adapted to relatively small temperature variations, small changes in the extremes can have relatively large impacts, such as alteration of ecosystems and species extinction.*
>
> *The ratio of extreme precipitation expressed by very wet days (R95p) to the total wet-day precipitation (PRCPTOT) represents the annual contribution of very wet days to the total annual wet-day precipitation (R95PTOT), which is relevant for societal impacts (Sillmann et al. 2013). The CDO based R95p computes the percentage of wet days (PR ≥ 1 mm) with daily precipitation amount greater than the 95th percentile of all wet days during the climate base reference period: 1976-2005. This index based on a percentile threshold takes into account the respective precipitation climatologies of different regions. The simple daily intensity index (SDII) describes the daily precipitation amount averaged over all wet days in a year. The maximum 5 day precipitation index (RX5day) describes the monthly or annual maximum of 5 day precipitation accumulations. This index is often used to describe changes in potential flood risks as heavy rain conditions over several consecutive days can contribute to flood conditions. The consecutive dry-day index (CDD) represents the length of the longest period of consecutive dry days (i.e., days with PR<1 mm) in a year ending in that year. If a dry spell does not end in a particular year and spans a period longer than 1 year (as may happen in very dry regions), then CDD is not reported for that year and the accumulated dry days are carried forward to the year when the spell ends.*

## 3.1 Temperature Extremes

The observations based on IMD daily gridded data over India since 1951 indicate that there is evidence of changes in some climate-related extremes. Figure 2.8 shows that there is an overall decrease in the number of cold nights (TN10p) and cold days (TX10p), and an increase in the number of warm nights (TN90p) and warm days (TX90p). There is a consistent decrease in cold nights (TN10p) and cold days (TX10p) from the late 20th to the 21st century in all RCP scenarios (Figures 2.8a and 2.8b). The mean decrease is generally more pronounced for TN10p than for TX10p. TN10p decreases to near 0% and TX10p to 2% by the end of 21$^{st}$ century for RCP4.5 scenario. Thus there will be virtually no cold nights or days over India as defined for the 1976–2005 reference base periods under the future projections. The spread among the RCMs (shading in Fig. 2.8) generally becomes smaller as the projection approaches the zero exceedance rates as more models simulate fewer cold nights and days. The warm nights (TN90p) and warm days (TX90p) over India show a general increase in the exceedance rate toward the end of the 21st century (Figures 2.8c and 2.8d). The increase is more pronounced for TN90p than for TX90p. The mean increase in TN90p and TX90p for RCP8.5 scenario is

22

from about 10% in 1976–2005 to 80% and 65% by the end of 21st century, respectively. The smallest increases in TN90p and TX90p, to 25% and 22% respectively, occur for RCP2.6 scenario, followed by greater respective increases to 50% and 40% for RCP4.5 scenario.

The changes in the percentile indices based on minimum temperature (TN10p and TN90p) are more pronounced than those based on maximum temperature (TX10p and TX90p). The largest decreases in TN10p and largest increases in TN90p projected over India are typical for tropical regions that are characterized by small day-to-day temperature variability so that changes in mean temperature are associated with comparatively larger changes in exceedance rates below the 10th and above the 90th percentiles.

The spatial pattern of the projected multi-RCM ensemble mean of the annual frequency of temperature indices shows that more rapid decreases in cold nights (TN10p) and cold days (TX10p) are expected along the west coast and the adjoining peninsular region than in other parts of India by mid 21st century for RCP4.5 and RCP8.5 scenarios (Fig. 2.9a and 2.9b). The increase in annual frequency of warm nights (TN90p) and warm days (TX90p) are also projected to be higher over these same regions by mid 21st century for RCP4.5 and RCP8.5 scenarios (Fig. 2.9c and 2.9d).

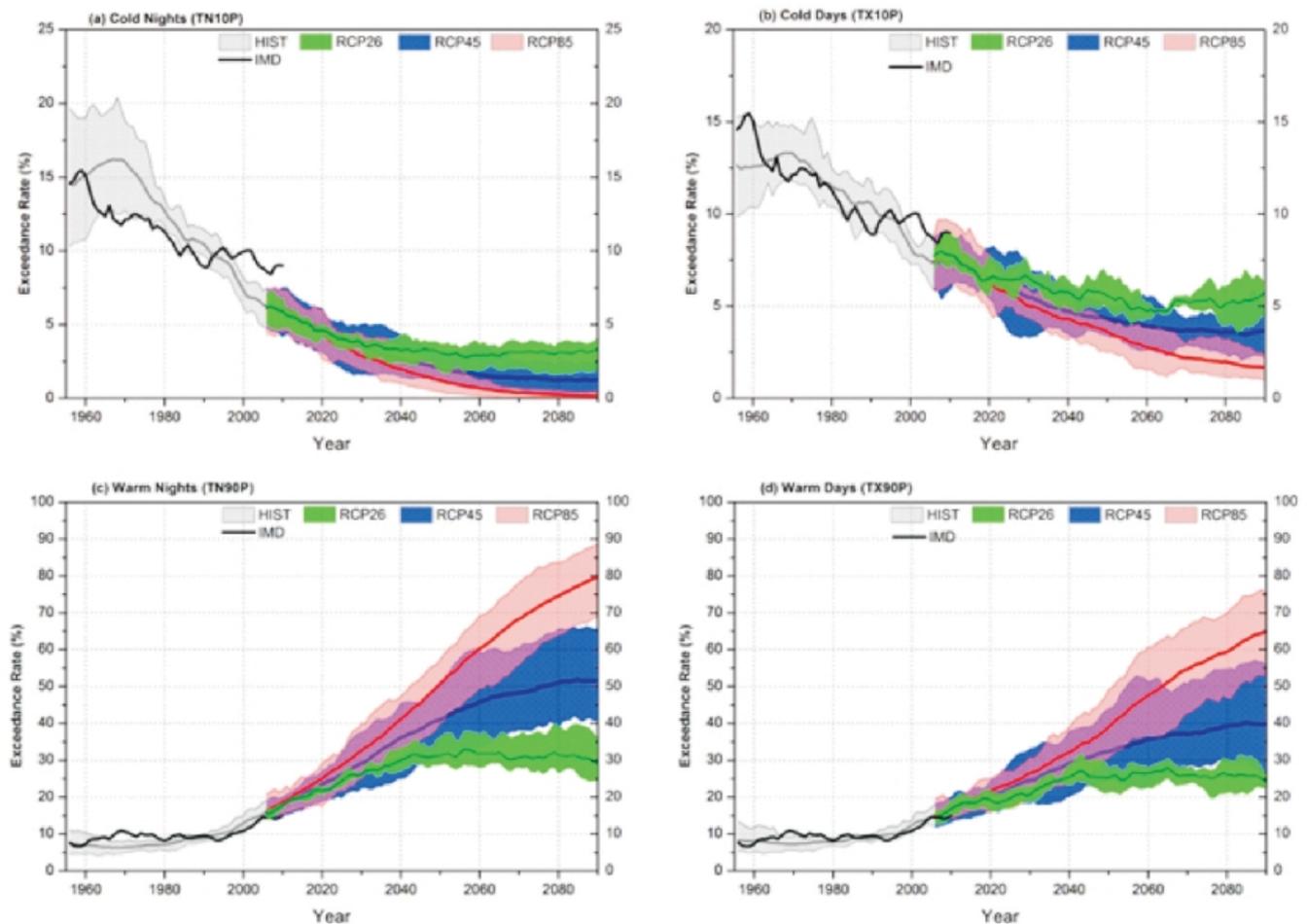

*Figure 2.8* *India averages of temperature indices over land as simulated by the CORDEX South Asia multi-RCM ensemble (see Table 2.1) for the RCP2.6 (green), RCP4.5 (blue), and RCP8.5 (red) displayed for the percentile indices (a) cold nights (TN10p), (b) cold days (TX10p), (c) warm nights (TN90p), and (d) warm days (TX90p). Changes are displayed as absolute exceedance rates (in %). By construction the exceedance rate averages to about 10% over the base period 1976-2005. Solid lines show the ensemble mean and the shading indicates the range among the individual RCM. The black line show the observed indices based on IMD gridded data. Time series are smoothed with a 20 year running mean filter.*

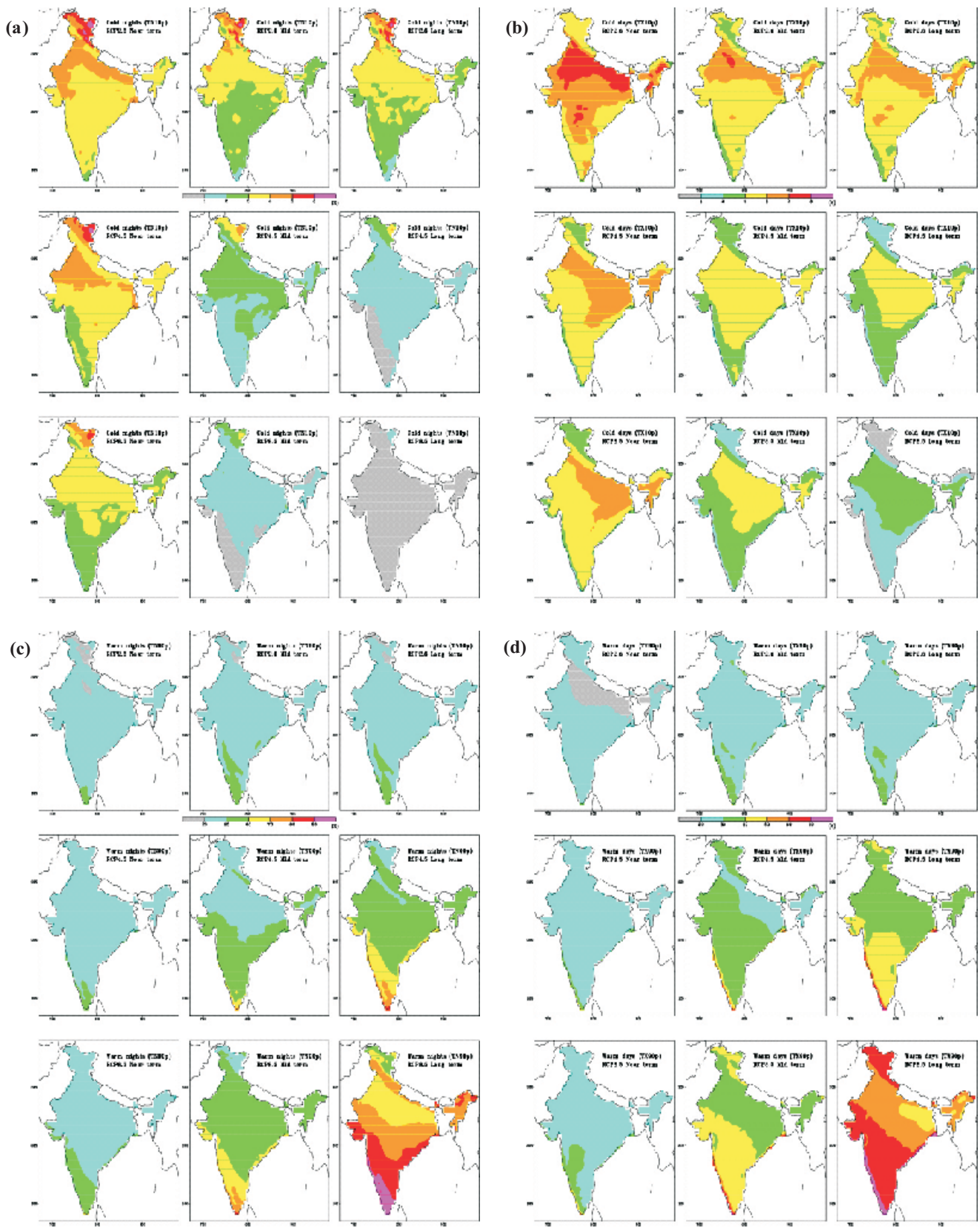

*Figure 2.9* CORDEX South Asia multi-RCM ensemble mean of the annual frequency of (a) cold nights (TN10p), (b) cold days (TX10p), (c) warm nights (TN90p), and (d) warm days (TX90p), temporally averaged for near-term (2016-2045), mid-term (2036–2065) and long-term (2066–2095) climate under RCP2.6, RCP4.5 and RCP8.5 scenarios, as absolute values of the exceedance rate (in %). By construction the exceedance rate averages to about 10% over the base period 1976–2005.

## 3.2 Precipitation Extremes

Changes in the selected precipitation indices (see details in Box 2.1) relative to the 1976–2005 reference period are expressed in percentage terms. Indian land averaged precipitation indices areprojected to increase in the 21st century (Fig. 2.10). Relative increases in maximum 5-day precipitation (RX5day; Fig. 2.10c), which represents a more extreme aspect of the precipitation distribution, are greaterover time than those for the contribution of very wet days to total wet day precipitation (R95PTOT; Fig. 2.10a) and the simple daily intensity index (SDII; Fig. 2.10b). In RCP8.5, R95PTOT and SDII are projected toincrease by 15% and 21%, respectively, by year 2100, whereasRX5day is projected to increase by 38%. However the spread among the CORDEX South Asia individual RCMs (shading in Fig. 2.10) is highest for RCP8.5 scenario, and remain overlapping with RCP4.5 scenario throughout the 21st century for these precipitation extreme indices. The all India averaged temporal evolution of maximumconsecutive dry days (CDD) is not shown here as the temporaland spatial variability of this index over Indian land area is verylarge.

The spatial pattern of the projected multi-RCM ensemble mean of the precipitationextreme indices shows that relatively higher increase in the contribution of very wet days to total wet day precipitation (R95PTOT; Fig. 2.11a), the daily intensity (SDII; Fig. 2.11b), and in the maximum 5-day precipitation (RX5day; Fig. 2.11c) are expectedalong the west coast and the adjoining peninsular region than in other parts of India by mid $21^{st}$ century for RCP4.5 and RCP8.5 scenarios. The striping plotted in Fig. 2.11 indicates where more than 70% of the RCM realizations concur on an increase (vertical) or decrease (horizontal) in theextreme precipitation indices for the RCP scenarios.The striping is not plotted for R95PTOT as the RCM realizations have more than 70% consensus for increase over entire India for the three RCP scenarios. Although the maximum number of consecutive dry days (CDD; Fig. 2.11d) is projected to increase over many parts of India, the RCM consensus is generally found only in parts of the Indian peninsular region throughout the $21^{st}$ century for RCP8.5 scenario. The increases in CDD combined with increases in RX5day indicates an intensification of both dry and wet seasons along the west coast and the adjoining peninsular region over India.

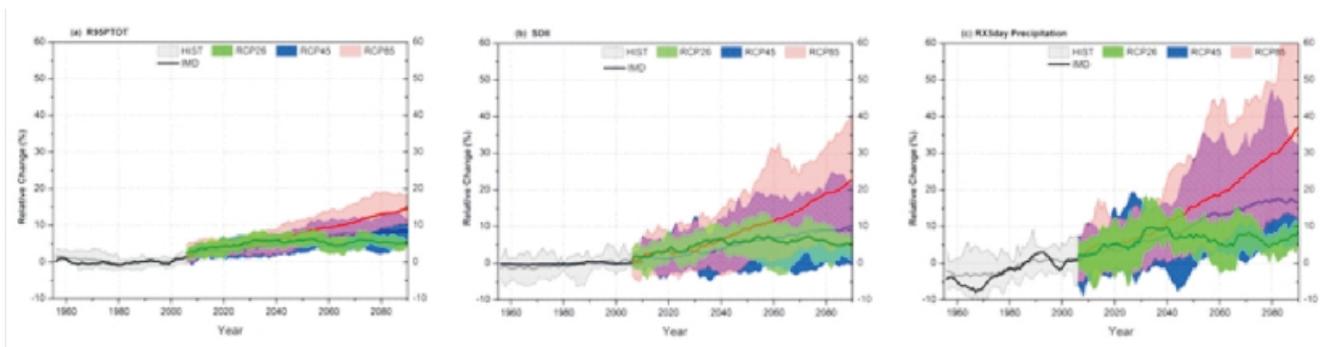

*Figure 2.10 India averages of precipitation indices over land as simulated by the CORDEX South Asia multi-RCM ensemble (see Table 1) for the RCP2.6 (green), RCP4.5 (blue), and RCP8.5 (red) displayed for the absolute indices (a) contribution of very wet days to total wet day precipitation (R95PTOT), (b) maximum 5-day precipitation (RX5day), (c) simple daily intensity index (SDII), and (d) maximum number of consecutive dry days (CDD). Changes are displayed relative to the reference period 1976-2005 (in %). Solid lines show the ensemble mean and the shading indicates the range among the individual RCMs. Time series are smoothed with a 20 year running mean filter.*

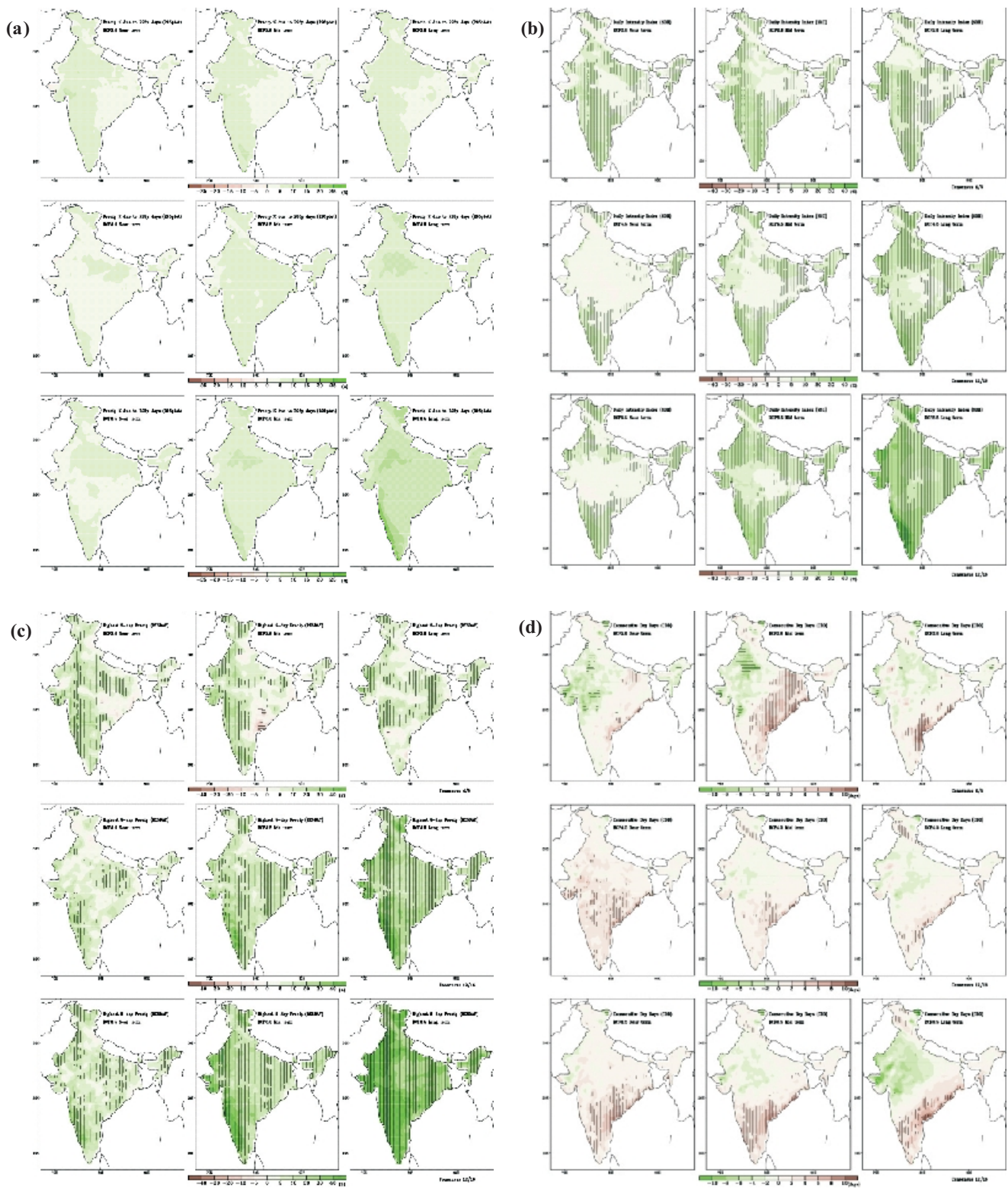

*Figure 2.11* CORDEX South Asia multi-RCM ensemble mean for the absolute precipitation indices (a) contribution of very wet days to total wet day precipitation (R95PTOT), (b) simple daily intensity index (SDII), (c) maximum 5-day precipitation (RX5day), and (d) maximum number of consecutive dry days (CDD), temporally averaged for near-term (2016-2045), mid-term (2036–2065) and long-term (2066–2095) climate under RCP2.6, RCP4.5 and RCP8.5 scenarios, displayed as changes relative to the reference period 1976-2005 (in %). Striping indicates where at least 70% of the RCM realizations concur on an increase (vertical) or decrease (horizontal) in the future scenarios. The striping is not plotted for R95PTOT as the RCM realizations have more than 70% consensus for increase over India for the three RCP scenarios.

## 4. Acknowledgments

The IITM-RegCM4 simulations were performed using the IITM Aaditya high power computing resources. The Director, IITM is gratefully acknowledged for extending full support to carry out this research work. IITM receives full support from the Ministry ofEarth Sciences, Government of India. The World Climate Research Programme's Working Group onRegional Climate, and the Working Group on Coupled Modelling, former coordinatingbody of CORDEX and responsible panel for CMIP5 are sincerely acknowledged. The climatemodeling groups (listed in Table 2.1) are sincerely thanked for producing and making availabletheir model output. The Earth System Grid Federationinfrastructure (ESGF; http://esgf.llnl.gov/index.html) is also acknowledged. The Climate Data Operatorssoftware (CDO; https://code.zmaw.de/projects/cdo/) and the Grid Analysis and Display System (GrADS; http://iges.org/grads/) were extensively used throughout this analysis.
## 5. References

Giorgi, F., Coppola, E., Solmon, F., et al. (2012), RegCM4:model description and preliminary tests over multiple CORDEX domains. Clim. Res., 52, 7–29, doi: https://doi.org/10.3354/cr01018.

INCCA(2010),Climate Change and India: A 4X4 Assesment - A sectoral and regional analysis for 2030s. http://www.envfor.nic.in/division/indian-network-climate-change-assessment

IPCC (2012), Managing the Risks of Extreme Events and Disasters toAdvance Climate Change Adaptation. A Special Report of Working Groups I and II of the Intergovernmental Panel on Climate Change, Cambridge University Press, Cambridge, UK, and New York, NY, USA, 582 pp.

NATCOMM2 (2012) IndiaSecond National Communication tothe United Nations Framework Conventionon Climate Change.http://unfccc.int/resource/docs/natc/indnc2.pdf

Samuelsson, P., Jones, C. G., Willen, U., et al. (2011), The Rossby Centre Regional Climate model RCA3: model description and performance. Tellus, 63A, 4–23, doi: 10.1111/j.1600-0870.2010.00478.x.

Sanjay, J., Ramarao, M. V. S., Mujumdar, M., Krishnan, R.(2017), Regional climate change scenarios. Chapter of book: Observed Climate Variability and Change over the Indian Region. Editors: M. N. Rajeevan and ShaileshNayak, Springer Geology, 285–304, doi: 10.1007/978-981-10-2531-0.

Sengupta, A. and M. Rajeevan (2013), Uncertainty quantification and reliabilityanalysis of CMIP5 projections for the Indiansummer monsoon. Current Science, 105, 1692-1703.

Sillmann, J., V. V. Kharin, X. Zhang, F. W. Zwiers, and D. Bronaugh (2013), Climate extremes indicesin the CMIP5 multimodel ensemble: Part 1. Model evaluation in the present climate, J. Geophys.Res. Atmos., 118,1716–1733, doi:10.1002/jgrd.50203.

Zhang, X., L. Alexander, G. C. Hegerl, P. Jones, A. K. Tank, T. C. Peterson,B. Trewin, and F. W. Zwiers (2011), Indices for monitoring changes inextremes based on daily temperature and precipitation data, WIREs Clim.Chang., 2, 851–870, doi:10.1002/wcc.147.
27